# Analysis of Large-Scale Propagation Models for Mobile Communications in Urban Area

M. A. Alim[*], M. M. Rahman, M. M. Hossain, A. Al-Nahid

Electronics and Communication Engineering Discipline,
Khulna University,
Khulna 9208, Bangladesh.

*Corresponding author.

*Abstract*— **Channel properties influence the development of wireless communication systems. Unlike wired channels that are stationary and predictable, radio channels are extremely random and don't offer easy analysis. A Radio Propagation Model (RPM), also known as the Radio Wave Propagation Model (RWPM), is an empirical mathematical formulation for the characterization of radio wave propagation as a function of frequency. In mobile radio systems, path loss models are necessary for proper planning, interference estimations, frequency assignments and cell parameters which are the basic for network planning process as well as Location Based Services (LBS) techniques. Propagation models that predict the mean signal strength for an arbitrary transmitter-receiver (T-R) separation distance which is useful in estimating the radio coverage area of a transmitter are called large-scale propagation models, since they characterize signal strength over large T-R separation distances. In this paper, the large-scale propagation performance of Okumura, Hata, and Lee models has been compared varying Mobile Station (MS) antenna height, Transmitter-Receiver (T-R) distance and Base Station (BS) antenna height, considering the system to operate at 900 MHz. Through the MATLAB simulation it is turned out that the Okumura model shows the better performance than that of the other large scale propagation models.**

*Keywords- Path Loss; Okumura model; Hata model; Lee model;*

## I. INTRODUCTION

In mobile radio systems, path loss models are necessary for proper planning, interference estimations, frequency assignments, and cell parameters which are basic for network planning process as well as LBS techniques that are not based on GPS system [3]. A Radio Propagation Model (RPM), also known as the Radio Wave Propagation Model (RWPM) or the Radio Frequency Propagation Model (RFPM), is an empirical mathematical formulation for the characterization of radio wave propagation as a function of frequency, distance and other conditions. Most radio propagation models are derived using a combination of analytical and empirical methods. In general, most cellular radio systems operate in urban areas where there is no direct line-of-sight path between the transmitter and receiver and where the presence of high rise buildings causes severe diffraction loss.

Propagation models that predict the mean signal strength for an arbitrary transmitter-receiver seperation distance are useful in estimating the radio coverage area of a transmitter and are called large-scale propagation model. On the other hand, propagation models that characterize the rapid fluctuations of the received signal strength over very short travel distances or short time durations are called small scale or fading models [1].

In this paper, the wideband propagation performance of Okumura, Hata, and Lee models has been compared varying Mobile Station (MS) antenna height, propagation distance, and Base Station (BS) antenna height considering the system to operate at 900 MHz. Through the MATLAB simulation it turned out that the Lee model outperforms the other large scale propagation models.

## II. LITERATURE REVIEW

Path loss characteristics of a channel are commonly important in wireless communications and signal propagation. Path loss may occur due to many effects, such as free-space loss, refraction, diffraction, reflection, aperture-medium coupling loss and absorption. Path loss is also influenced by terrain contours, environment (urban or rural, vegetation and foliage), propagation medium (dry or moist air), the distance between the transmitter and the receiver, and the height of antennas [4].

Path loss normally includes propagation losses caused by

- The natural expansion of the radio wave front in free space (which usually takes the shape of an ever-increasing sphere)

- Absorption losses (sometimes called penetration losses)

- When the signal passes through media not transparent to electromagnetic waves, diffraction losses when part of the radiowave front is obstructed by an opaque obstacle and

- Losses caused by other phenomena.

The signal radiated by a transmitter may also travel along many and different paths to a receiver simultaneously; this effect is called multipath. Multipath can either increase or decrease received signal strength, depending on whether the individual multipath wavefronts interfere constructively or destructively.

In wireless communications, path loss can be represented by the path loss exponent, whose value is normally in the range of 2 to 4 (where 2 is for propagation in free space, 4 is for relatively lossy environments. In some environments, such as buildings, stadiums and other indoor environments, the path loss exponent can reach values in the range of 4 to 6. On the other hand, a tunnel may act as a waveguide, resulting in a path loss exponent less than 2 [4].

The free-space path loss is denoted by $L_p(d)$, which is

$$\overline{L}_p(d) = -20\log_{10}\left(\frac{c/f_c}{4\pi d}\right) \text{ (dB)}$$

where, c = velocity of light, fc = carrier frequency and d = distance between transmitter and receiver [2].

For log-distance path loss with shadowing the path loss is denoted by $\overline{L}_p(d)$, which is

$$\overline{L}_p(d) \propto \left(\frac{d}{d_0}\right)^n, \ d \ge d_0$$

or equivalently,

$$\overline{L}_p(d) = \overline{L}_p(d_0) + 10n\log_{10}\left(\frac{d}{d_0}\right) \text{ (dB)}, \ d \ge d_0$$

where, n = path loss component, d0 = the close-tin reference distance (typically 1 km for macrocells, 100m for microcells), d = distance between transmitter and receiver [2].

## III. MATERIALS AND METHODS

Calculation of the path loss is usually called prediction. Exact prediction is possible only for simpler cases, such as the above-mentioned free space propagation or the flat-earth model. For practical cases the path loss is calculated using a variety of approximations.

Statistical methods (also called stochastic or empirical) are based on fitting curves with analytical expressions that recreate a set of measured data. Among the most commonly used such methods are Okumura Model, Hata Model, and Lee's Model.

In the cities the density of people is high. So the more accurate loss prediction model will be a good help for the BTS mapping for optimum network design. Among the

Radio Propagation Models (RPM) city models are to be analysed in this paper to find the best fitting city model. The well known propagation models for urban areas are:

i) Okumura Model

ii) Hata Model

iii) Lee's Model

### A. Okumura Model

The Okumura model for urban areas is a Radio propagation model that was built using the data collected in the city of Tokyo, Japan. This model is applicable for frequencies in the range of 150 MHz to 1920 MHz and distances of 1 km to 100 km. It can be used for the base stations antenna heights ranging from 30m to 100m.

To determine path loss using Okumara's model, the free space path loss between the points of interest is first determined and then the value of $A_{mu}(f,d)$ is added to it along with correction factors according to the type of terrain. The expression of the model [1]:

$$L_{50}(dB) = L_F + A_{mu}(f,d) - G(h_{te}) - G(h_{re}) - G_{AREA} \quad \ldots\ldots (1)$$

where, $L_{50}$ = The 50th percentile (i.e. median) value of propagation path loss., $L_F$ = The Free Space propagation Loss in dB, $A_{mu}$ = Median attenuation.relative to free space in dB, $G(h_{te})$ = The base station antenna height gain factor, $G(h_{re})$ = Mobile antenna height gain factor and $G_{AREA}$ = The gain due to the type of environment.

$$G(h_{te}) = 20\log_{10}\left(\frac{h_{te}}{200}\right); \quad 100m \rangle h_{te} \rangle 30m$$

$$G(h_{re}) = 10\log_{10}\left(\frac{h_{re}}{3}\right); \quad h_{re} \le 3m$$

$$G(h_{re}) = 20\log_{10}\left(\frac{h_{re}}{3}\right); \quad 10m \rangle h_{re} \rangle 3m$$

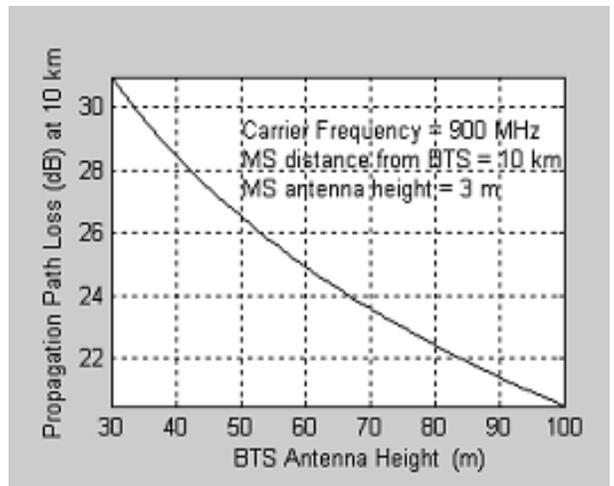

Figure 1. Propagation Path Loss due to the change in BTS antenna height for Okumura model.

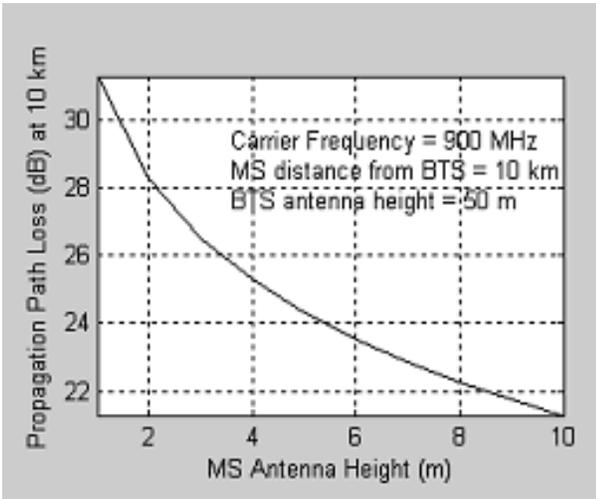

Figure 2.   Propagation Path Loss due to the change in MS antenna height for Okumura model.

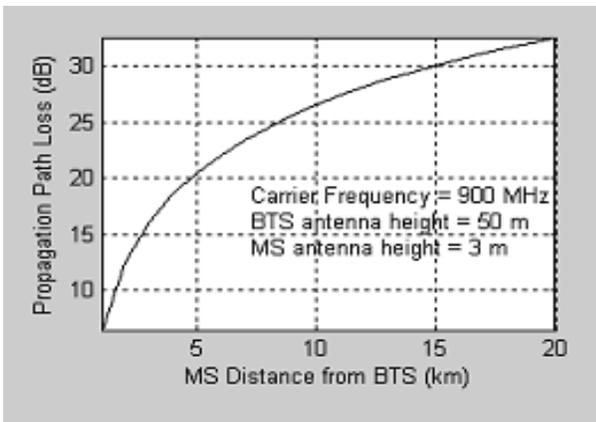

Figure 3.   Propagation Path Loss due to T-R separation for Okumura model.

## B.  Hata Model

Hata Model is based on the Okumara's model where some correction factors are included. It works in the frequencies range from 150 MHz to 1500 MHz.. The standard formula for median path loss in urban areas is given by [1].

$$L_{50}(urban)(dB) = 69.55 + 26.16\log_{10} f_c - 13.82\log_{10} h_{te}$$
$$- a(h_{re}) + (44.9 - 6.55\log_{10} h_{te})\log_{10} d \quad \ldots\ldots\ldots(2)$$

where, fc =The frequency from 150 MHz to 1500 MHz, hte = The effective base station antenna height (30m to 200m), hre = The effective mobile antenna height (1m to 10m), d = The trnasmitter-receiver (T-R) distance in km and a(hre) = The correction factor for effective mobile antenna hieght which is a function of the size of coverage area. For a large city it is given by,

$$a(h_{re}) = 8.29(\log_{10} 1.54 h_{re})^2 - 1.1 dB$$

for  $f_c \leq 300 MH_Z$

$$a(h_{re}) = 3.2(\log_{10} 11.75 h_{re})^2 - 4.97 dB$$

for  $f_c \geq 300 MH_Z$

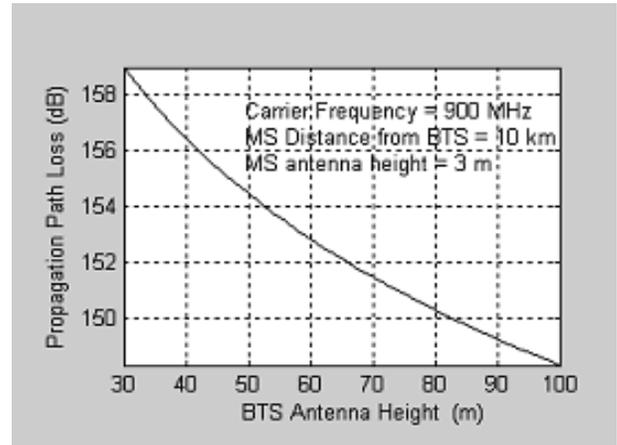

Figure 4.   Propagation Path Loss due to the change in BTS antenna height for Hata model.

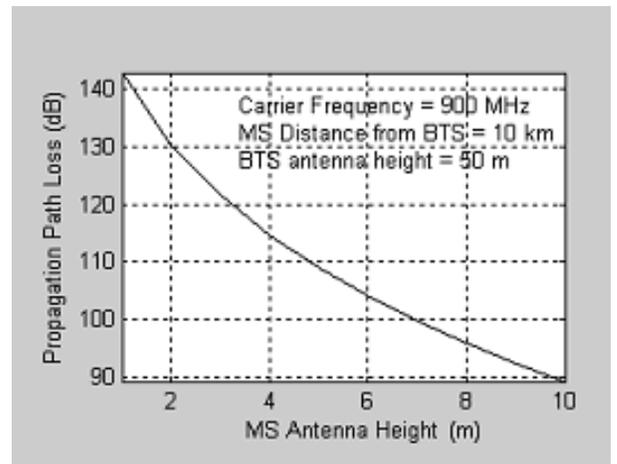

Figure 5.   Propagation Path Loss due to the change in MS antenna height for Hata model.

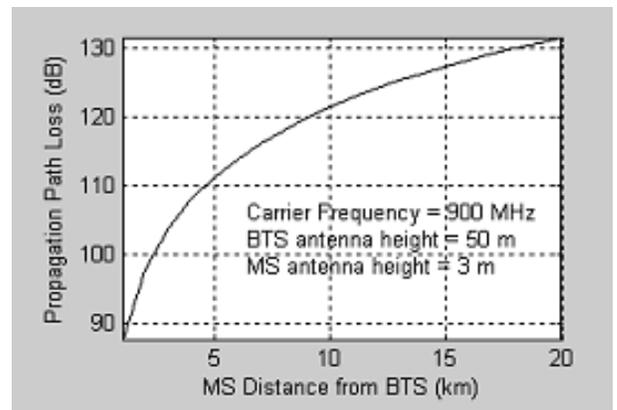

Figure 6.   Propagation Path Loss due to T-R separation for Hata model.

*C.  Lee Model*

Lee's path loss model is based on empirical data chosen so as to model a flat terrain. Large errors arise when the model is applied to a non-terrain. However, Lee's model has been known to be more of a "North American model" than that of Hata.

The propagation loss calculated as:

$$L(dBm) = 124 + 30.5\log_{10}\left(\frac{d}{d_0}\right) + 10k\log_{10}\left(\frac{f}{f_c}\right) - \alpha_0 \quad\text{................(3)}$$

where, d is in km, f and fc is in MHz, k = 2 for fc < 450 MHz and in suburban/open area and 3 for fc > 450 MHz and in urban area, d0 = 1.6 km.

In this analysis the parameter values taken for calculations are:

Carrier Frequency fc = 900 MHz

Nominal (Calibration) Distance do = 1.6 km

Base Mobile Station Antenna hb = 30.48 m

Mobile Station Antenna Height hm = 3 m

Base Station Transmit Power Pb = 10 W

Base Station Antenna Gain Gb = 6 dB

Mobile Station Antenna Gain Gm = 0 dB with respect to isotropic antenna.

*f* is the transmitted frequency, d is the Transmitter-Receiver distance and α0 is a correction factor to account for BS and MS antenna heights, transmit powers and antenna gains that differ from the nominal values. As such, when the prevailing conditions differ from the nominal ones, then α0 is given by:

$$\alpha_0 = 10\log_{10}(\alpha_1\alpha_2\alpha_3\alpha_4\alpha_5)$$

where:

α1 = (new BTS antenna height (m) / 30.48 m)2

α2 = (new MS antenna height (m) / 3 m)v ;for MS antenna height < 3, v =1 and for MS antenna height >3, v =2;

α3 = (new transmitter power / 10 W)2, in this paper the value of α3 is taken 1.

α4 = new BS antenna gain correction factor = (Gb / 4), Here α4 is considered 6 dB.

α5 = frequency correction factor = (*f*/*f*c)-n ; for 2<n<3 and f and *f*c is in MHz. *f*c is taken 900 MHz.

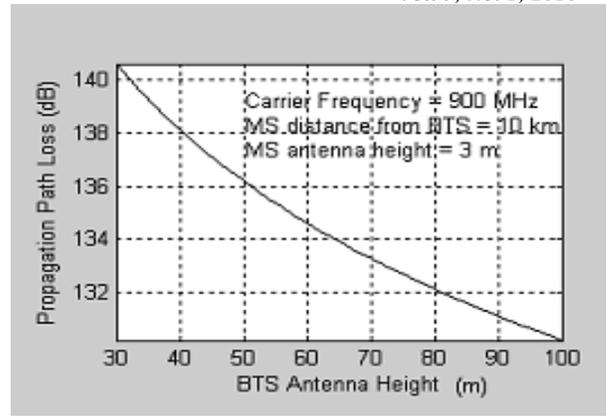

Figure 7.   Propagation Path Loss due to the change in BTS antenna height for Lee model.

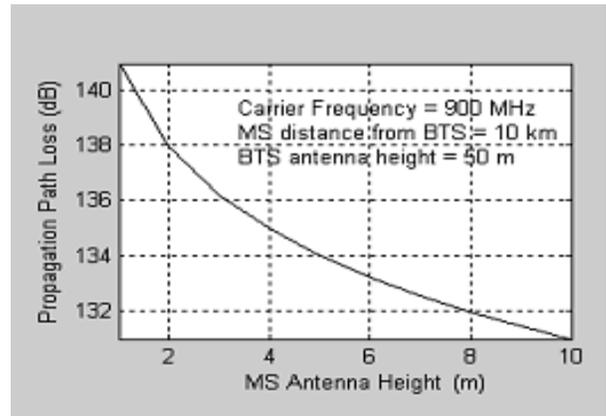

Figure 8.   Propagation Path Loss due to the change in MS antenna height for Lee model.

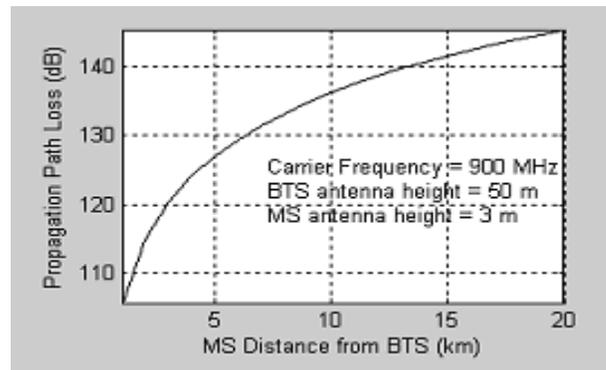

Figure 9.   Propagation Path Loss due to the T-R separation for Lee Model

## IV.   RESULT

From Fig. 10, it is seen that the propagation path loss decreases due to the increase in BTS antenna height for all the models. For Hata model the loss is maximum, for Lee model the loss is medium and for Okumura model the loss is minimum. From Fig. 11, it is seen that the propagation path loss increases with the decrease in MS antenna height for all the models. For Lee model the loss is maximum, for Hata model the loss is medium and for Okumura model the loss is minimum.

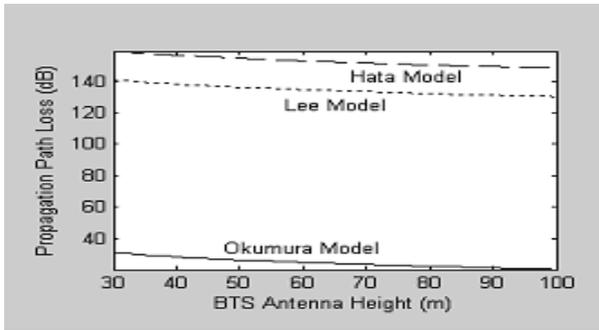

Figure 10. Comparison of Propagation Path Loss due to the change in BTS antenna height.

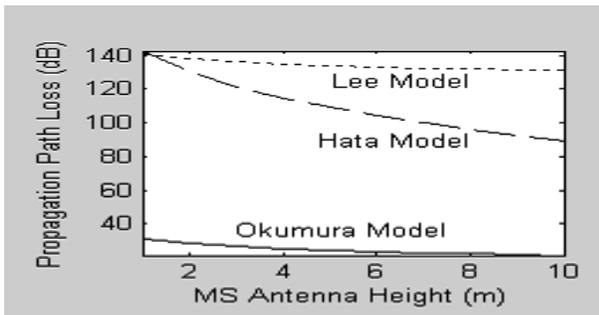

Figure 11. Comparison of Propagation Path Loss due to the change in MS antenna height.

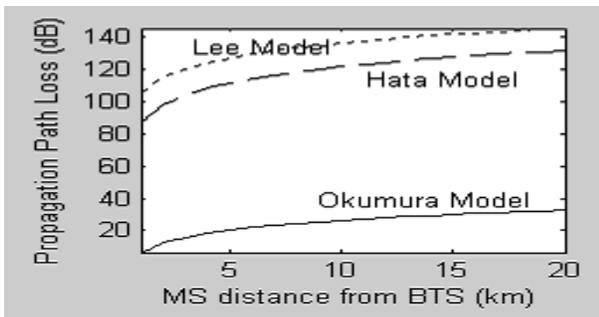

Figure 12. Comparison of Propagation Path Loss due to T-R separation in BTS antenna height.

From Fig. 12 it is seen that for Lee model the propagation path loss is highest due to the increase in MS distance from BTS than the other two models and Okumura model has the lowest path loss. From the analyses it is seen that overall Okumura model shows the better performance than that of the other two models.

## V. CONCLUSION

In this paper, three widely known large scale propagation models are studied and analyzed. The analyses and simulation was done to find out the path loss by varying the BTS antenna height, MS antenna height, and the T-R separation. Okumura model was seen to represent low power loss levels in the curves. The result of this analysis will help the network designers to choose the proper model in the field applications. Further up-gradation in this result can be possible for the higher range of carrier frequency.